\documentclass{aa}

\usepackage{graphicx}
\usepackage{txfonts}

\usepackage{caption}
\usepackage{subcaption}
\begin{document}

   \title{Probing initial distributions of orbital eccentricity and disc misalignment via polar discs}


   \author{
   Simone Ceppi\inst{1,3}\thanks{simone.ceppi@unimi.it},
   Nicolás Cuello\inst{2},
   Giuseppe Lodato\inst{1},
   Cristiano Longarini\inst{1,3},
   Daniel J. Price\inst{3},
   Daniel Elsender\inst{4},\and
   Matthew R. Bate\inst{4}
          }

   \institute{Dipartimento di Fisica, Università degli Studi di Milano, via Celoria 16, 20133 Milano, Italy
             \and
             Univ. Grenoble Alpes, CNRS, IPAG, F-38000 Grenoble, France
             \and
             School of Physics and Astronomy, Monash University, Clayton, VIC 3800, Australia
             \and
             Department of Physics and Astronomy, University of Exeter, Stocker Road, Exeter EX4 4QL, UK
             }

      \date{Accepted January 2024}

\abstract
{In a population of multiple protostellar systems with discs, the sub-population of circumbinary discs whose orbital plane is highly misaligned with respect to the binary's orbital plane constrains the initial distribution of orbital parameters of the whole population.
We show that by measuring the polar disc fraction and the average orbital eccentricity in the polar discs, one can constrain the distributions of initial eccentricity and mutual inclination in multiple stellar systems at birth.}

    \titlerunning{Protobinary properties via polar discs}
    \authorrunning{Ceppi et al.}

   \maketitle
%

\section{Introduction}
    \label{sec:pa}

    The tilt of an accretion disc orbiting an eccentric binary has two alternative equilibrium configurations. If the mutual inclination (initial disc tilt with respect to the stellar orbital plane) is below a critical angle for polar alignment, \citep{Farago&Laskar10, Zanazzi&Lai18, Cuello&Giuppone19} the disc is expected to nodally precess, and align to the stellar orbital plane due to viscous dissipation \citep{Bate+00, Lubow&Ogilvie00}.
    If the initial mutual inclination is higher than the critical angle for polar alignment, \cite{Aly+15} and \cite{Martin+17} showed that the circumbinary disc angular momentum vector precesses around the binary orbit eccentricity vector. Due to viscous dissipation, the disc evolves to a polar configuration in which the mutual inclination is around 90$^\circ$ (depending on disc parameters, see \citealt{Martin+19}). 

    \cite{Ceppi+23} showed that, in the general case of a disc in a multiple stellar system, if the disc is orbiting more than two stars the polar alignment mechanism is highly suppressed\footnote{Polar alignment could still occur under the right conditions (e.g. very small semi-major axis ratios or for very small discs). \cite{Ceppi+23} derived an analytical criterion to assess the stability of the polar configuration in triples (see also the criterion obtained by \citealt{Lepp+23}).}. Conversely, if the disc is orbiting a pair of stars with additional bodies outside the disc, polar alignment is at least as likely as in the pure circumbinary disc case.

    The properties of multiple stellar systems are already used to gain insights into the physics driving stellar and planet formation evolution. The available statistics on multiple stellar systems supports that the vast majority of stars are born in a multiple stellar system and that stellar multiplicity increases with stellar mass \citep{Larson72, Duchene&Kraus13, Offner+2022}. These are crucial constraints for numerical experiments on the collapse of molecular clouds \citep[][]{Bate+02, Krumholz+12, Bate18, Bate19,Mathew+21, Mathew+23, Lebreuilly+23}. Orbital parameter distributions, such as the semi-major axis distribution \citep{Duquennoy&Mayor91, Raghavan+10}, or the binary mass ratio distribution \citep{Moe&DiStefano17, El-Badry+19} also help to constrain formation mechanisms. The presence of more than two stars is expected to change the shape of these distributions, encoding additional information about the dependence on multiplicity \citep{Smith+97, Ceppi+22, Offner+2022}. 

    In this work, we show that, from the current distribution of mutual misalignment and eccentricity --- such as the one in \citet{Czekala+19} --- and with increased statistics, we are able to constrain the distributions of eccentricity and inclination between orbital planes and discs at birth. Such distributions, in turn, depend on which physical properties are more significant in the formation process of these objects. In addition to dynamical interactions between stars \citep[][]{Bate+02, Bate18, Elsender+23}{}{}, the presence and the strength of magnetic fields \citep{Price+07, Wurster+19, Zhao+20, Lebreuilly+21}, different metallicities \citep{Elsender&Bate21} and the level of turbulence in the cloud \citep{Bate+10, Walch+12} may play a significant role in setting the initial properties of disc and star populations.
    
    The paper is organised as follows: 
    In Section~\ref{sec:toymodel} we present a model to compute the fraction of polar discs and the mean eccentricity of stellar orbits hosting polar discs. By applying this method to a population of hierarchical systems and of pure binaries, we derive the relationship between the initial and evolved distributions of system properties. Finally, in Section~\ref{sec:3cases} we constrain the initial conditions for both populations.
    In Section~\ref{sec:syntpop} we discuss our results and we give our conclusions in Section~\ref{sec:concl}.

\section{Initial conditions and polar disc population}\label{sec:toymodel}
    \subsection{Polar disc fraction and mean polar systems eccentricity}\label{sec:gen-model}

       As soon as the condition for polar alignment is satisfied, an accretion disc starts oscillating around the polar configuration. The disc dissipates the oscillation on a fraction of the viscous timescale \citep{Lubow+18, Zanazzi&Lai18}. Thus, it is reasonable to assume that all discs able to go polar in the initial population will do so. Then, if we neglect the impact of subsequent external interactions, the fraction of polarly aligned discs we observe in a given evolved population is directly linked to the initial conditions (eccentricity and misalignment distributions) in a forming population. 

        In this section, we build a toy model to estimate the expected fraction of polar discs and their eccentricity distribution in an evolved young stellar population (Class II). 
        Given the observed and theoretically predicted preference for low mutual inclinations (e.g. \citet{Czekala+19, Elsender+23}), we describe the initial distribution of mutual inclination with a normalised exponential distribution:
        \begin{equation}\label{eq:initpdfb}
            P_\beta(\beta) = \frac{1}{N_\beta}\exp\left(-\frac{\beta}{\sigma_\beta}\right),
        \end{equation}
        where $\beta$ is the mutual inclination, $\sigma_\beta$ is a parameter that regulates the shape of the distribution and $N_\beta$ normalises the distribution over the support considered, i.e. from $\beta=0$ to $\pi/2$.

        The initial distribution of eccentricity is described by a normalised power law distribution:
        \begin{equation}\label{eq:initpdfe}
        P_{\rm e}(e) = \frac{1}{N_{\rm e}}e^\alpha,
        \end{equation}
        where $e$ is the orbital eccentricity, $\alpha$ is a parameter regulating the distribution shape and $N_{\rm e}$ normalises the distribution over $e=0$ to~$1$. This is in line with surveys of eccentricities \citep{Duquennoy&Mayor91, Raghavan+10, Hwang+22} and it allows for a thermal distribution ($P(e) \propto e$) that can be produced by repeated N-body interactions \citep{Jeans19, Ambartsumian37, Heggie75}.

        Thus, the two-dimensional probability density function for the initial condition is given by:
        \begin{equation}
            P(e,\beta)=P_{\rm e}(e)P_{\beta}(\beta).
        \end{equation}
        
        For an orbit co-rotating with the central system, the critical angle for polar alignment is\footnote{for a counterrotating orbit $\beta_{\rm crit}(e_{\rm b}, \Omega) = \pi- \arcsin{\sqrt{\frac{1-e_{\rm b}^2}{1-5e_{\rm b}^2\cos{\Omega}^2+4e_{\rm b}^2}}}$} \citep[][]{Farago&Laskar10, Zanazzi&Lai18, Cuello&Giuppone19}
        \begin{equation}
            \beta_{\rm crit}(e, \Omega) = \arcsin{\sqrt{\frac{1-e^2}{1-5e^2\cos{\Omega}^2+4e^2}}},
            \label{eq:betacrit}
        \end{equation}
        where $e$ is the orbital eccentricity and $\Omega$ is the disc longitude of the ascending node. 
        
        Supposing that all discs above $\beta_{\rm crit}$ go polar and neglecting subsequent external interactions, we can integrate over the initial mutual inclination above $\beta_{\rm crit}$, to obtain the distribution of polar discs as a function of system eccentricity  $P_{\rm pol}(e)$. At this stage, we consider $\Omega=\pi/2$ and we will fix $P_{\rm pol}(e)$ to take into account a distribution of longitude of the ascending node in Sec.~\ref{sec:acc-long}. Thus,
        \begin{equation}\label{eq:Ppol}
            P_{\rm pol}(e) = \int_{\beta_{\rm crit}(e,\frac{\pi}{2})}^{\pi/2}P(e,\beta) {\rm d}\beta.
        \end{equation}

        Integrating $P_{\rm pol}$ over the eccentricity we obtain the expected fraction of polar discs ($F_{\rm p}$) in an evolved population, that is the ratio between the number of polar discs and the total number of discs in the population:
        \begin{equation}\label{eq:3PF}
            F_{\rm p} = \int_{0}^{1}P_{\rm pol}(e) {\rm d}e.
        \end{equation}

        Additionally, from $P_{\rm pol}$ we can compute the mean eccentricity of stellar systems hosting polar discs, i.e.: 
        \begin{equation}\label{eq:3avee}
            \langle e \rangle = \int_{0}^{1}eP_{\rm pol}(e) {\rm d}e.
        \end{equation}

        Integrals in Eqs.~\eqref{eq:3PF} and~\eqref{eq:3avee} are challenging to solve due to the dependencies of $\beta_{\rm crit}$. In the next section we present two assumptions to take into account a distribution of longitude of the ascending nodes in the case of hierarchical systems and pure binaries.

        \subsection{Taking into account the longitude of ascending node}\label{sec:acc-long}
        \begin{figure}
            \centering
    	    \includegraphics[width=0.4\textwidth]{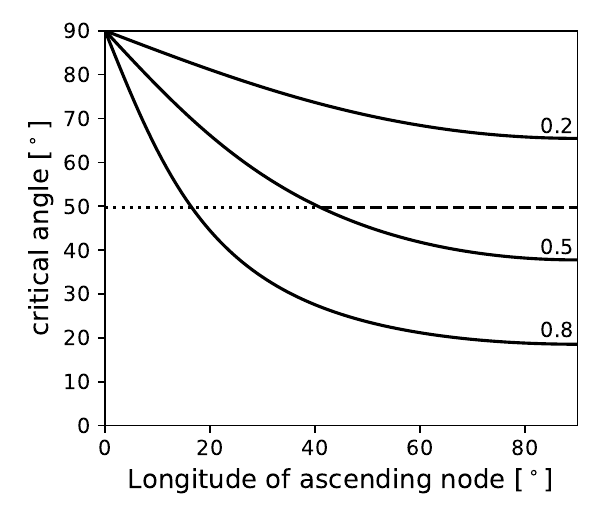}
            \caption{Critical angle for polar alignment as a function of the disc longitude of the ascending node for a fixed binary eccentricity ($e=0.2, 0.5, 0.8$). Given a mutual inclination (e.g. $50^\circ$) and an eccentricity (e.g. $e=0.5$), the dashed (dotted) line is the $\Omega$ interval in which the disc will go perpendicular (coplanar) to the orbital plane.} 
            \label{fig:Omegabin}
        \end{figure}

        Figure~\ref{fig:Omegabin} shows how the critical angle $\beta_{\rm crit}$ in Eq.~\eqref{eq:betacrit} depends on the longitude of the ascending node ($\Omega$) for different orbital eccentricities. Let us take the subpopulation of circumbinary discs with a given mutual misalignment $\beta_{\rm sp}$ orbiting pure binaries with a given orbital eccentricity $e_{\rm sp}$. We have a $\Omega_{\rm crit}$ so that the critical angle for polar alignment $\beta_{\rm crit}(e_{\rm sp}, \Omega_{\rm crit})=\beta_{\rm sp}$. All discs with $\Omega<\Omega_{\rm crit}$ (dotted line in Figure~\ref{fig:Omegabin}) will not polar align. All discs with a $\Omega>\Omega_{\rm crit}$ (dashed line) will polar align.
        
        If we suppose a uniformly distributed $\Omega$, the fraction of discs that will polar align is the ratio between the length of the dashed curve and the width of the $\Omega$ interval (i.e. $\pi/2$).
        Thus, the fraction $f$ of discs that will polarly align is:
        \begin{equation}
            f(e_{\rm sp}, \beta_{\rm sp}) = 1 - \frac{2}{\pi}\Omega_{\rm crit}(e_{\rm sp},
            \beta_{\rm sp}).
        \label{eq:f}
        \end{equation}
       
        In the case of a binary population, we have to take into account the factor $f(e, \beta)$. To compute the distribution of polar discs with respect to orbital eccentricity, Equation~\eqref{eq:Ppol} becomes:
        \begin{equation}\label{eq:2Ppol}
            P_{\rm pol}(e) = \int_{\beta_{\rm crit}(e)}^{\pi/2}P(e,\beta)f(e, \beta) {\rm d}\beta    .                        
        \end{equation}
        
        In circumbinary discs in hierarchical systems, the precession of the eccentricity vector drives the polar alignment process \citep{Ceppi+23}. The orbit precesses on a shorter timescale than the timescale for coplanar alignment. Hence, precession could lower the critical angle for polar alignment to its minimum value because the system quickly explores the $\Omega$ for which the critical angle is minimum. This is true only if, while the system is exploring different longitudes of the ascending node, the tilt of the disc does not decrease significantly. Otherwise, the polar alignment of the disc would still be favoured compared to binary systems but the initial longitude of the ascending node would nevertheless be relevant. If we suppose this hypothesis to hold, for discs orbiting the inner binary of a triple what really matters is the minimum critical angle no matter the longitude of ascending node. Independent of $\Omega$, if the disc inclination is higher than the minimum critical angle (the one for $\Omega=90^\circ$) the disc will polar-align. Therefore, for triples $\beta_{\rm crit}(e, \Omega) = \beta_{\rm crit}(e, \Omega=90^\circ).$ Thus, Eq.~\eqref{eq:Ppol} can be written as
        \begin{equation}
            \begin{split}\label{eq:3Ppol}
            P_{\rm pol}(e) &= \int_{\beta_{\rm crit}(e)}^{\pi/2}P(e,\beta) {\rm d}\beta =\\
                           &= -P_{\rm e}(e)\frac{\sigma_\beta}{N_\beta}
                           \left[\exp\left(-\frac{\pi/2}{\sigma_\beta}\right)-
                                 \exp\left(-\frac{\beta_{\rm crit}(e)}{\sigma_\beta}\right)\right].
           \end{split} 
        \end{equation}

         With the previous assumptions, we are left only with the eccentricity dependence both for binaries and hierarchical systems.

        \subsection{Polar fraction and mean eccentricity for multiple systems}\label{sec:application}

        \begin{figure*}
            \centering
         \begin{subfigure}[b]{0.9\textwidth} 
    	    \includegraphics[width=\textwidth]{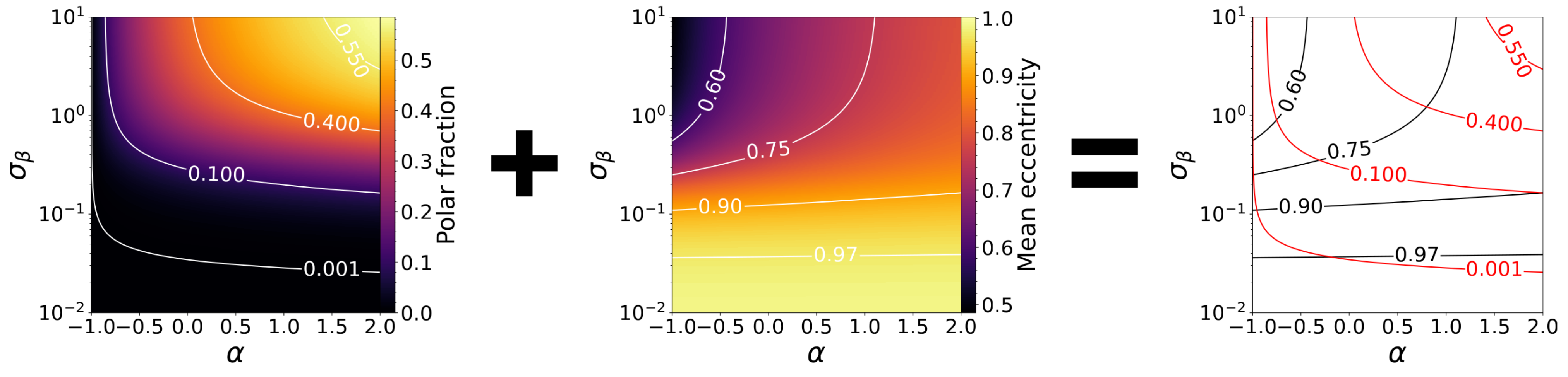}

            \label{fig:binoper}
            \end{subfigure}
            \hfill
             \begin{subfigure}[b]{0.9\textwidth} 
            \includegraphics[width=\textwidth]{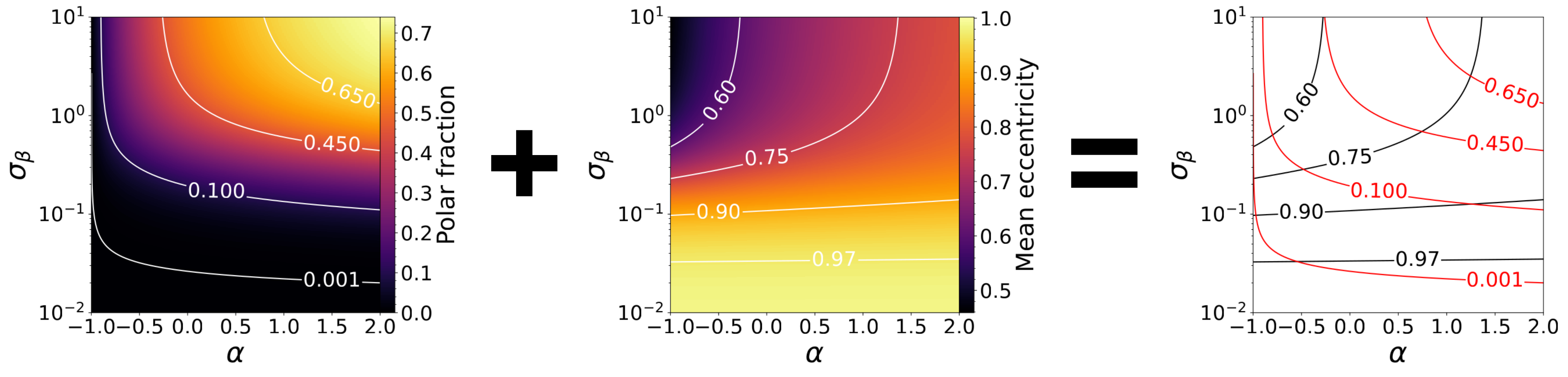}
            \label{fig:trioper}
            \end{subfigure}
            \caption{Fraction of polar discs ($F_{\rm p}$, leftmost column) and mean orbital eccentricity of systems hosting polar discs ($\langle e\rangle$, central column) for binary system population (top row) and triple system population (bottom row). As shown by the rightmost column, each point of the $\alpha-\sigma_\beta$ parameter space is uniquely characterized by a pair of $F_{\rm p}$ and $\langle e \rangle$ (crossing of two contour lines, red contour for polar fraction, black contour for mean eccentricity). } \label{fig:oper}
        \end{figure*}

        Given a pair of values $\alpha$ and $\sigma_\beta$ --- the parameters of the distributions of eccentricity and mutual angle, respectively --- we are now able to compute the expected polar disc fraction of an evolved population and the mean eccentricity of systems hosting a polar disc with Equations~\eqref{eq:3PF} and~\eqref{eq:3avee}, respectively. For pure binaries and hierarchical systems, the polar disc distributions are given by Eqs.~\eqref{eq:2Ppol} and~\eqref{eq:3Ppol}, respectively.

        We semi-analytically computed these integrals over the $\alpha$--$\sigma_\beta$ parameter space for the initial distributions. The parameter $\alpha$ ranges between $-1$ and $2$, while $\sigma_\beta$ ranges between $0.01$ and $+\infty$. The left panels in Figure~\ref{fig:oper} show how the fraction of expected polar discs in an evolved population $F_{\rm p}$ depends on $\alpha$ and $\sigma_\beta$ for binary and hierarchical systems. In general, the binary population is less prone to host polar discs compared to systems with more than two stars. This is due to the different $\beta_{\rm crit}$ we used to describe the two populations. The higher the probability of having high eccentricity or mutual inclination, the more likely it is to find configurations with a high mutual inclination and orbital eccentricity which go polar more easily. Thus, $F_{\rm p}$ increases for higher $\alpha$ or $\sigma_\beta$.

        The central panels in Figure~\ref{fig:oper} show the mean eccentricity of orbits hosting polar discs ($\langle e \rangle$) over the $\alpha$--$\sigma_\beta$ parameter space for binaries and hierarchical systems. The binary polar population tends to have higher $\langle e \rangle$, again due to the $f$ factor (Eq.~\eqref{eq:f}). Indeed, the $\Omega$ interval allowing polar alignment is larger for more eccentric systems. Thus, we have more polar discs around highly eccentric systems. 
        Lowering $\sigma_\beta$ raises $\langle e \rangle$. Indeed, higher eccentricity values are required in regions where systems are mildly misaligned. Conversely, lowering $\alpha$ means lowering $\langle e \rangle$. The less likely is to have very eccentric systems in the initial population, the less eccentric the polar population will be.

        Each pair of $\alpha$--$\sigma_\beta$ uniquely connects to a pair of $F_{\rm p}-\langle e \rangle$. This can be seen with contour lines in the right panel of Figure~\ref{fig:oper}. A given pair of polar fraction $F_{\rm p}$ and mean eccentricity $\langle e \rangle$ contour lines cross only at one point of the $\alpha$--$\sigma_\beta$ parameter space. Thus, we are able to numerically invert the $\alpha$--$\sigma_\beta$ and $F_{\rm p}-\langle e \rangle$ coordinate systems to obtain plots in Figure~\ref{fig:sigmas}. This plot showcases how $\alpha$ and $\sigma_\beta$ depend on $F_{\rm p}$ and $\langle e \rangle$. Doing so, once we have constrained $F_{\rm p}$ and $\langle e \rangle$ from observations in a population we can constrain $\alpha$ and $\sigma_\beta$ in the initial condition.

\subsection{Measurement of polar fraction and mean eccentricity}\label{sec:observations}

Statistics on mutual inclinations between discs and stellar orbital planes are fairly scarce at the moment. Even if it is relatively easy to measure the disc inclination with respect to the sky plane, constraining the stellar orbital inclination is challenging. Thus, in the most recent literature surveys \citep{Czekala+19, Zurlo+23} the sample size is around 15 discs. Among these discs, only one accretion disc is confirmed in a polar configuration: HD~98800B \citep[][]{Kennedy+19,Zuniga-Fernandez+21}. Additionally, two discs are likely polar\footnote{There is a degeneracy of 180$^\circ$ in the orbital longitude of the ascending node \citep{Czekala+19}.} (HD~142527, \citet{Balmer+22}, and SR~24  N, \citet{Fernandez-Lopez+17}).
Another notable example is 99 Her, which is a polar debris disc that likely evolved from a polar accretion disc \citep[][]{Smallwood+20}{}{}. HD~98800B and SR~24N have additional companions outside the circumbinary disc. The semimajor axis ratios for the two systems are about 0.02 and 0.01 for HD~98800 and SR 24, respectively. The presence of additional bodies could possibly affect the process of polar alignment, for example, exciting Kozai-Lidov oscillations in the disc. However, the outer orbits of these triples are too wide to drastically impact the inner disc \citep{Martin+14}.

In the following, we show that the two different prescriptions for $\Omega$ in binaries and hierarchical systems with more than two stars result in similar values for the mean eccentricity, and in minor differences in terms of polar fraction distributions compared to the current uncertainties we have in surveys. Thus, here we do not distinguish between binaries and hierarchical systems in the sample.

We estimate the mean eccentricity of systems hosting polar discs averaging over confirmed polar systems (HD98800B), systems that are likely polar (SR24 N and HD142527) and the polar debris disc 99 Her. We obtain an average eccentricity of $\langle e \rangle^{\rm obs}=0.67 \pm 0.11$.

To evaluate the polar disc fraction, if we take into account the single confirmed polar system (HD98800B) we end up with a polar fraction of $0.08$. If, however, we include also the additional two likely polar systems (SR24 N and HD142527) the fraction raises up to $0.25$. We take the average of these two values as a very rough estimate for the polar fraction, with their standard deviation as an error. The result is $F_{\rm p}^{\rm obs}=0.17 \pm 0.08$. This fraction should be considered as an upper limit. In a sample of well-measured discs and orbital planes inclinations, there is a bias towards systems in which those quantities of interest are measured properly (for example because they are promising polar discs). 

In the following, we use $F_{\rm p}^{\rm obs}$ and $\langle e \rangle^{\rm obs}$ as references for exploring the parameter space of the initial conditions.

\section{Mapping observations onto the parameter space}\label{sec:3cases}
\begin{figure*}
         \centering
         \begin{subfigure}[b]{0.4\textwidth} 
             \includegraphics[width=\textwidth]{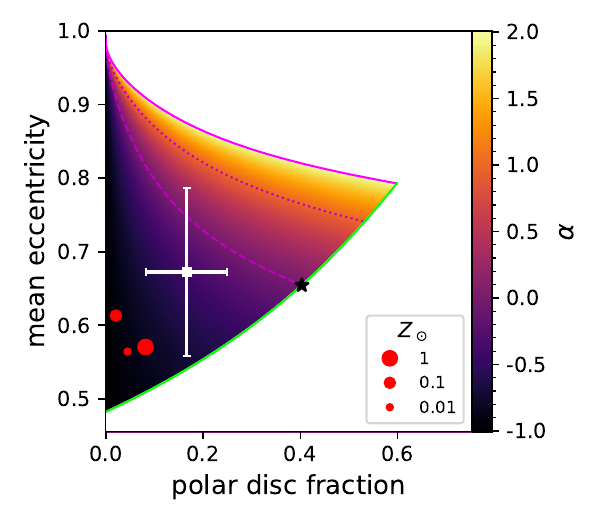}
             \caption{}
             \label{fig:2sige}
         \end{subfigure}
         \hfill
         \begin{subfigure}[b]{0.4\textwidth}
             \centering
             \includegraphics[width=\textwidth]{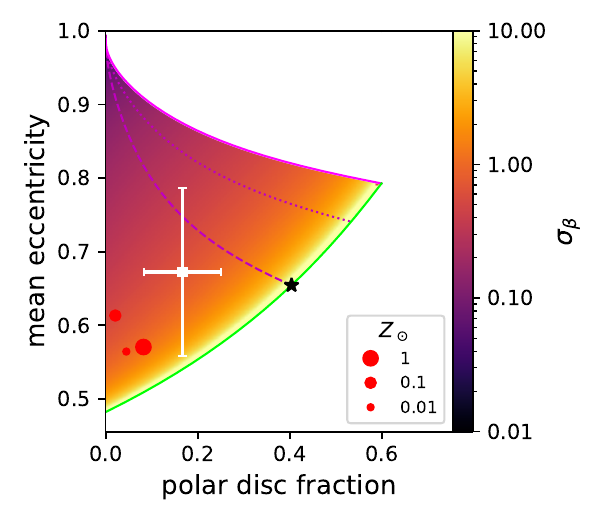}
             \caption{}
             \label{fig:2sigb}
         \end{subfigure}
         \begin{subfigure}[b]{0.4\textwidth} 
             \centering
             \includegraphics[width=\textwidth]{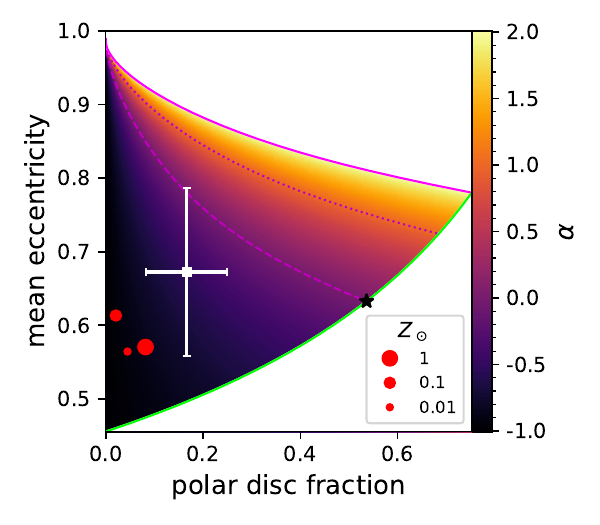}
             \caption{}
             \label{fig:3sige}
         \end{subfigure}
         \hfill
         \begin{subfigure}[b]{0.4\textwidth}
             \centering
             \includegraphics[width=\textwidth]{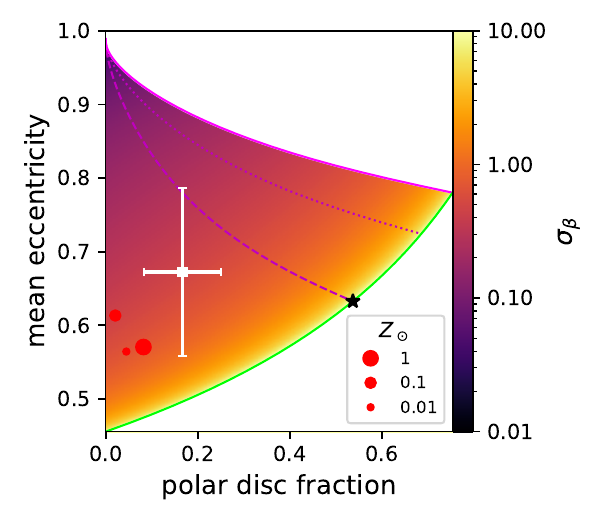}
             \caption{}
             \label{fig:3sigb}
         \end{subfigure}
         \caption{$\alpha$ (left) and $\sigma_\beta$ (right) as a function of $\langle e \rangle$ and $F_{\rm p}$ for binary (top) and triple (bottom) populations. White regions are $F_{\rm p}-\langle e\rangle$ pairs that cannot be generated by any $\alpha-\sigma_\beta$ pair. Purple curves are the parameter space region in which $\alpha=2,1,0$ for solid, dotted and dashed, respectively. Green curves are the parameter space region in which $\sigma_\beta=\infty$. Black star is the solution for randomly distributed mutual inclination and eccentricity. White box point with error bars is the $F_{\rm p}^{\rm obs}$ and $\langle e\rangle^{\rm obs}$ observation values derived in section \ref{sec:observations}. Red circle points are $F_{\rm p}-\langle e\rangle$ pairs measured in the four \citet{Bate19} molecular cloud collapse simulations with different metallicities.  }
         \label{fig:sigmas}
    \end{figure*}

For a binary (triple) population, Figures~\ref{fig:2sige} and~\ref{fig:2sigb} (\ref{fig:3sige} and~\ref{fig:3sigb}) show $\alpha$ and $\sigma_\beta$, respectively, as a function of $\langle e \rangle$ and $F_{\rm p}$. 

White regions are regions where no $\alpha$ and $\sigma_\beta$ pair results in the given pair of $\langle e \rangle$ and $F_{\rm p}$. In the following, we check which initial distribution is more likely to form the observed polar-aligned disc population in binaries and triples.

    \subsection{Randomly distributed initial condition}\label{sec:ran-ran}
    The first assumption we test is flat distributions in both orbital eccentricity and mutual inclination. This corresponds to taking the limit for $\sigma_\beta$ approaching infinity and $\alpha=0$ in Eqs.~\eqref{eq:initpdfb} and \eqref{eq:initpdfe} respectively. The assumptions are: i) star formation at the molecular cloud level gives no preferential orbital eccentricity; ii) there is no preferred mutual inclination at the onset of stellar and disc formation.
    
    Under these assumptions, we can compute analytically both the polar fraction and the mean eccentricity of the triple polar population, solving Eqs.~\eqref{eq:3PF} and~\eqref{eq:3avee} respectively with $P_{\rm e}=1$ and $P_\beta=2/\pi$. The results are $F^{\rm tri}_{\rm p}=0.54$ and $\langle e \rangle^{\rm tri}=0.63$.
    
    For the binary population, we numerically integrate Eq.~\eqref{eq:2Ppol} since the $f$ factor makes the integral not analytically solvable. Results for binaries are $F^{\rm bin}_{\rm p}=0.40$ and $\langle e \rangle^{\rm bin}=0.65$. We find these points in Fig.~\ref{fig:sigmas} at the crossing of the green and the dashed purple curves, marked with a star. Indeed, green and dashed purple curves are the limit for $\sigma_\beta$ approaching infinity and $\alpha=0$, respectively. In surveys, we find a polar fraction $F_{\rm p}^{\rm obs}=0.17 \pm 0.08$ and a mean eccentricity  $\langle e \rangle^{\rm obs}=0.67 \pm 0.11$ (see Section~\ref{sec:observations}). Even if the mean eccentricity in this configuration is compatible with the observed one, the observed polar fraction is not compatible with the expected polar fraction for a binary/triple population. 

    These circumbinary discs form around young stars due to accretion of surrounding gas and circularisation. Assuming that there is no correlation between the forming system's and the disc' angular momenta, then rather than assuming a flat distribution of mutual inclination, we expect $P_\beta=\sin{\beta}$. In this case, we can take advantage of Eq.~(16) in \citet{Aly+15} which gives the fraction of configurations undergoing polar precession for a randomly distributed $\Omega$, thus applicable to the binary population. Integrating over the eccentricity we obtain an expected $F_{\rm p}^{\rm bin}=1-\tanh{(2/\sqrt{5})}/\pi\approx0.54>0.4$. This estimate is higher than the one previously computed given that, in this case, higher inclinations are favoured compared to coplanar configurations. Likewise, for the triple population, we analytically solve Eq.~\eqref{eq:3PF} with $P_\beta=\sin{\beta}$, obtaining $F_{\rm p}^{\rm tri}=(5-\sqrt{5})/4\approx0.69>0.54$.

    This result implies that non-uniform distributions --- either in mutual inclination and/or orbital eccentricity --- are needed to explain the observed polar population (or that many polar discs have been missed, which seems unlikely). 

    \subsection{Correlated orbit-disc mutual inclinations}
    The first hypothesis we relax is the random initial distribution of mutual misalignment.  We are still bound to move along the dashed purple curve in Figures~\ref{fig:2sigb} and~\ref{fig:3sigb} (where $\alpha=0$). Over this restricted parameter space region, $\langle e \rangle$ has a lower limit given by the $\sigma_\beta=\infty$ case (i.e. 0.65 and 0.63 for binaries and triples, respectively), while $F_{\rm p}$ can span values from $0$ to the $\sigma_\beta=\infty$ case (i.e. 0.40 and 0.54 for binaries and triples, respectively). The restricted parameter space region is compatible with $\langle e \rangle^{\rm obs}$ and $F_{\rm p}^{\rm obs}$. In particular, observations constrain $\sigma_\beta$ to range between $0.25$ and $0.78$ for binaries and between $0.17$ and $0.43$ for triples which implies a narrow $\beta$ distribution around $\beta=0$, i.e. close to aligned orbit-disc systems should be more common at birth. Such correlation between the angular momenta of the disc and of the binary/triple orbit is needed to describe the observed polar population, under the assumption of randomly distributed initial orbital eccentricities.

    \subsection{Non-flat initial conditions}
    We now allow $\alpha$ and $\sigma_\beta$ to explore the whole parameter space to fully exploit the information contained in the measurement of $\langle e \rangle$ and $F_{\rm p}$.
    The observed mean eccentricity and polar fraction select a region of possible values of $\alpha$ and $\sigma_\beta$ highlighted by the error bars in Fig.~\ref{fig:sigmas}. For binaries, we obtain $\alpha\leq 0.6$ and $0.26\leq\sigma_\beta\leq1.8$. As for the triples, $\alpha\leq0.46$ and $0.23\leq\sigma_\beta\leq1.87$. 
    
    Again, the small statistics suggest the presence of a correlation between the angular momenta of the discs and stellar systems. As for the eccentricity, even if it is still marginally compatible with a random distribution, present data suggest a decreasing function ($\alpha<0$). Indeed, with a flat or slightly increasing eccentricity distribution, we would expect an higher mean eccentricity or an higher polar fraction.

\section{Discussion}
\label{sec:syntpop}
\subsection{Constraining initial conditions in multiple stellar systems}
Our analysis suggests that, to be compatible with present data about polar discs, initial distributions both for mutual inclinations and eccentricities have to be non-uniform. While we expect the angular momenta of forming discs and multiple systems to be correlated, the distribution of eccentricities we find does not completely match with observational results from surveys of evolved multiple stellar systems.

The observed evolved eccentricity distribution ranges from a uniform distribution ($\alpha=0$) for orbits with semi-major axis of the order of 100~AU \citep{Raghavan+10}, up to an (increasing) thermal distribution ($\alpha=1$) for 500~AU and becomes even steeper for larger systems \citep{Duquennoy&Mayor91, Tokovinin20, Hwang+22}. Conversely, we find an $\alpha$ between $-1$ and $0.6$, pointing towards a slightly decreasing distribution, only marginally compatible with a uniform one. 

First, we have to consider that we do not observe systems hosting circum-multiple discs with semi-major axis larger than about 100~AU \citep[e.g.][]{Czekala+19} nor numerical simulations form them \citep[e.g.][]{Elsender+23}, thus we are interested in comparing our results with this range of semi-major axis in which $\alpha$ is closer to our findings (systems of $\sim200$~AU have $\alpha\sim0.6$). In addition, gravitational interactions tend to raise the average eccentricity of the population. However, the thermalisation of the distribution could be only partially explained with gravitational interaction during cluster evolution which should act for much longer in order to push $\alpha$ from 0 to 1 \citep{Heggie75, Weinberg+87}. Thus, depending on the thermalisation timescale, we could reconcile our findings with observed distributions. 

In addition, the orbital eccentricity can evolve through interactions between the binary and the circumbinary disc. For example, \citet{DOrazio&Duffell21} numerically show that the hydrodynamical interaction with a coplanar disc could lead to preferred values of binary eccentricity. Even if it is not clear if this result is compatible with eccentricity surveys, the fact that our model does not take into account that the eccentricity of observed systems could be systematically different from the initial condition is also something to investigate in the future. To the best of our knowledge, there are no studies regarding this effect on misaligned circumbinary discs.

\subsection{Comparison with cluster formation simulations}    
    Nowadays, it is possible to perform detailed numerical simulations of the collapse of molecular clouds. In such simulations, the initial conditions of the cloud are set (e.g. amount of turbulence, strength of magnetic fields, metallicity), and regulate the distribution of the parameters of the population of the forming protosystems. By tuning the properties of the cloud, one could aim at reproducing the measured distribution of inclination and eccentricity. This would result in an indirect measurement of the molecular cloud properties. 

    In this work, we applied this analysis to the molecular cloud collapse simulations by \citet{Bate19}. We measured $\alpha$ and $\sigma_\beta$ in the newly born protostellar systems population by fitting the synthetic $P_{\rm pol}$ (Eq.~\eqref{eq:Ppol}) resulting from the simulation. We computed the expected $F_{\rm p}$ and $\langle e \rangle$ with Equations \eqref{eq:3PF} and \eqref{eq:3avee}, respectively. Finally, we compared the values of $F_{\rm p}$ and $\langle e \rangle$ obtained from the simulations with the ones measured in observations (see Section~\ref{sec:observations}). 
    
    The simulation set consists of four molecular clouds collapsing with four different metallicities (see details in \citealt{Elsender&Bate21}). We note that the higher the metallicity, the steeper the eccentricity distribution becomes. The steep decrease in eccentricity does not fit well with the same functional form as the observed distributions, on which we based our parameterisation. Thus, we are able to satisfactorily fit the three lower metallicities only. 

    The three red dots in Figure~\ref{fig:sigmas} represent the computed $F_{\rm p}$ and $\langle e \rangle$ for three different realisations of the same cloud but with different metallicities. The metallicity has an impact both on the eccentricity and mutual inclination distribution, hence the scatter in Figure~\ref{fig:sigmas}. However, it appears to have too little impact on the properties of the population.
    
    Regardless of the metallicity, the resulting polar disc fractions all are generally too low compared to observations. Given that the distribution of misalignment angles from the calculations of \citet{Bate19} and the observed systems are in good agreement \citep{Elsender+23}, this mismatch must be due to the lack of eccentric orbits in the simulations (possibly due to the simulations being too dissipative). We note however that, as discussed in Section~\ref{sec:observations}, the proposed observational value has to be considered as an upper limit. In other words, we cannot currently completely rule out such small polar disc fractions from observations.

    Finally, this analysis would benefit from numerical simulations producing distributions of mutual inclination and eccentricity, but for different sets of initial conditions (e.g. different amount of turbulence or magnetic field strength) --- such as \citet{Bate19} with metallicity. This would lead to more robust predictions and better constraints for the polar disc population. Provided accurate measurements of $F_{\rm p}$ and $\langle e \rangle$, the method illustrated here constitutes a powerful way to infer the initial conditions in molecular clouds from disc populations.

\section{Conclusions}
\label{sec:concl}
We showed how to measure the correlation between the inclination of accretion discs and of forming stellar multiple systems (i.e. the distribution of mutual inclination) and the distribution of orbital eccentricity of such stellar systems at the onset of star and disc formation. Using our model, we were able to compute the two fundamental parameters describing the initial distributions of disc-orbit mutual inclination ($\sigma_\beta$) and orbital eccentricity ($\alpha$). The only required measurements are the fraction of polar discs in a disc population ($F_{\rm p}$) and the mean eccentricity of systems hosting polar discs ($\langle e \rangle$). Despite the low statistics available, we find that:
\begin{enumerate}
    \item The observed disc population is not compatible with a randomly distributed initial distribution of mutual misalignment -- there must be a preference for aligned systems;
    \item The orbital eccentricity is marginally compatible with a random distribution as observed in field stellar systems with semi-major axis below $100$ au. The observed increasing eccentricity for wider orbits is still compatible with present data up to $\sim 200$ au. However, our model suggests a slight initial preference for circular orbits. We will investigate in future works if this discrepancy is compatible with the eccentricity evolution of young multiple stellar systems.
\end{enumerate}

The limitation of this proof-of-concept toy model lies in the simplified distributions taken to describe the initial conditions. This simplification allowed us to describe the distributions with only two parameters, facilitating the computation of the relations between $\alpha-\sigma_\beta$ and $F_{\rm p}-\langle e \rangle$ pairs, the parameter space discussion, and the comparison with data. This gives no degeneracy in the model and a strictly two dimensional parameter space. A better observational constraint on the polar disc fraction would improve the robustness of this method. Also, the impact of the interaction with a polar disc on the eccentricity of the binary system has to be investigated further.

In conclusion, by measuring the polar disc fraction and the distribution of mutual inclinations, we showed that it is possible to infer the initial eccentricity and mutual inclination distributions of binaries and triples at birth. This will shed light on formation processes within molecular clouds that affect the population of binary and triple stars.

\section*{Acknowledgements}
We thank the referee for the suggestions with which the manuscript substantially improved. This project received funding from the European Research Council (ERC) under the European Union Horizon Europe programme (grant agreement No. 101042275, project Stellar-MADE). DJP acknowledges Australian Research Council funding via DP220103767. SC's visit to Monash was funded by the EU Marie Curie RISE scheme DUSTBUSTERS (grant 823823).  The ideas discussed in this paper were partially formulated by SC, DJP, DE, and MRB while attending the Binary22 programme at the Kavli Institute of Theoretical Physics (KITP) at the University of California, Santa Barbara, which is supported in part by the National Science Foundation under Grant No. NSF PHY-1748958.

The data set consisting of the output from the calculations of Bate (2019) that are analysed in this paper is available from the University of Exeter’s Open Research Exeter (ORE) repository and can be accessed via the handle: http://hdl.handle.net/10871/35993.

%
%

\bibliographystyle{aa} 
\bibliography{example.bib} 

\end{document}